\documentstyle[epsfig]{aipproc}

\begin{document}
\title{Generating Non-Gaussian\\ Adiabatic Fluctuations\\ from
Inflation}

\author{D.S. Salopek\thanks{E-mail: dsscosmos@yahoo.com}
\thanks{Current mailing address: Space-Time Institute, 
14915-105 Avenue, Edmonton, Alberta, Canada T5P 4M2}}

\address{Department of Physics \& Astronomy\\
6224 Agricultural Road, University of British Columbia\\
Vancouver, Canada V6T 1Z1}

\address{To appear in}

\address{  
Proceedings of COSMO-98 \\ 
International Workshop on Particle Physics and the Early Universe\\
Monterey, California, Nov. 15-20, 1998 \\
Ed. D. Caldwell (Published by American Institute of Physics) }

\maketitle

\begin{abstract}
As the quality of cosmological data continue to improve,
it is natural to test the statistics
of primordial fluctuations: are they Gaussian or non-Gaussian?
I review a model which generates
non-Gaussian adiabatic fluctuations from inflation. 
Current investigations suggest that there
may possibly be a non-Gaussian signal in large angle cosmic
microwave background anisotropy data.
Statistics of microwave anisotropies could thus serve as a powerful 
probe of the very early Universe.
\end{abstract}

\section*{Introduction}
%
%
%
%

The inflationary expansion of the very early universe acts like a 
microscope that magnifies short scale quantum fluctuations to encompass
our entire observable Universe. In a sense, microwave background anisotropy
maps such as the COBE DMR maps provide an image of the vacuum state
that arose from inflation. If the field interactions during 
the inflationary epoch are linear, then one expects that the statistics
of vacuum fluctuations will be Gaussian --- this is the situation that most
researchers are hoping for.  However, there are indeed models where 
significant nonlinearities arise 
which would produce a non-Gaussian signal. Here, I will review such an
inflation model which produces non-Gaussian adiabatic fluctuations
\cite{S92}. 

By considering 3-point correlations of spherical harmonic amplitudes in 
COBE DMR maps,
Ferreira {\it et al} \cite{FMG98} have found evidence 
of a non-Gaussian signal (see also Magueijo \cite{M99}).    
They point out a non-Gaussian signal associated with the
$\ell = 16$ multi-pole of a spherical harmonic expansion. They are not 
able to associate the signal with
any known foreground emission, although they are concerned that 
the power takes a mysterious dip at this same value of 
$\ell$. Their work requires further investigation.       

The computation of nonlinear effects during inflation is a 
challenging problem. However, nonlinearities are essential to the 
formulation of the inflationary scenario.  For example, at the 
end of inflation, the inflaton must transfer 
its energy to a bath of radiation and matter; 
nonlinearities must thus be significant. The `heating' of the Universe
is an interesting problem that is currently under investigation
(see, {\it e.g.}, Kofman \cite{K99}). The computation of nonlinear effects 
during the slow-roll period of inflation ({\it e.g.}, before heating of the 
Universe) faces two severe theoretical obstacles: 

(1) The complete field theory that encompasses inflation is not known. 
String theory may provide the required framework
although this remains to be seen.

(2) Quantum gravity effects are important
during inflation. For example, it is generally agreed that the quantum
generation of tensor modes (gravitational waves) during inflation is a 
quantum gravity effect although it is only a `linear' effect. 

Soon after the COBE
detection in 1992, it was pointed out by Salopek \cite{SAL92} and 
Davis {\it et al} \cite{DHSST92} and subsequently verified by many others
that tensor modes from inflation could indeed contribute significantly to 
COBE's signal. In particular, Salopek \cite{SAL92} pointed out that at 
most 50\% of
COBE's signal could be attributed to tensor modes arising from 
power-law inflation otherwise there
would not be enough fluctuations to account for the observed clustering
of galaxies. (For a careful analysis of how current cosmological
data restricts tensor modes, consult Zibin {\it et al} \cite{ZSW99};
they confirm the above 50\% bound that was deduced over five years earlier.  
In this conference, Kinney \cite{K99} explains how the measurement of
polarization by future microwave background anisotropy experiments will help
to determine the contribution of tensor modes more accurately.) 

In order to bypass severe problems associated with nonlinear quantum 
effects of gravity, Stewart and Parry and I advocate solving the 
Hamilton-Jacobi equation for general relativity \cite{PSS94} 
which provides a nonlinear semiclassical description of gravity 
interacting with matter.  We attempt to imitate the
historical development of the theory of atomic spectra. Before
the development of the quantum theory in 1926, the semiclassical
theory of Bohr and Sommerfeld provided a useful although imperfect
description of various atoms. 

The non-Gaussian model described below
employs several unique features of the Hamilton-Jacobi equation.
It has a distinct observational signature:
it produces a non-Gaussian distribution for cosmic microwave
temperature anisotropies. This model is constrained by the analysis of 
DMR maps by
Kogut {\it et al} \cite{K96} and Ferreira {\it et al} \cite{FMG98}. 
Typically each non-Gaussian model 
produces a specific signature which allows one to discriminate 
among the different models. This property provides a powerful probe of
the very early Universe.  

\section*{Hamilton-Jacobi Equation for Cosmology}

In a Hamilton-Jacobi (HJ) formulation, there are just two laws that govern the
inflationary epoch:
\begin{equation}
{\cal H}(x) = 0 \, , \quad {\cal H}_i(x) = 0 \, . \label{CONSTRAINTS}
\end{equation}
The first legislates that the {\bf total energy density vanishes} 
at every point in the 
Universe, whereas the second legislates that the 
{\bf total momentum density vanishes} at every point. 
In fact, it is a great accomplishment
that one may write two mathematical laws that govern the very early Universe.
Explicitly, the energy and momentum densities may be written in terms of the 
scalar fields,
$\phi_a$, the 3-metric, $\gamma_{ij}$, and their respective momenta,
$\pi^{\phi_a}$, $\pi^{ij}$:
\begin{mathletters}
\begin{eqnarray}
{\cal H} =&&  
\kappa \, \gamma^{-1/2} 
\, [ 2 \gamma_{ik} \, \gamma_{jl} - \gamma_{ij} \gamma_{kl} ]
\pi^{ij} \, \pi^{kl} + 
{\kappa \over 2}  \sum_a \gamma^{-1/2} \left ( \pi^{\phi_a} \right )^2 
+ \kappa \gamma^{1/2} V(\phi_a) + \nonumber \\
&& \left \{ - { 1 \over 2 \kappa} \, \gamma^{1/2} \, R + 
{1 \over 2 \kappa} \, \gamma^{1/2} 
\,   \sum_a \phi_{a|i} \phi_{a}{}^{|i} \right \} \, , 
\label{ENERGY} \\
{\cal H}_i = && -2 \left ( \gamma_{ij} \, \pi^{jl} \right )_{,l} + 
\pi^{kl} \gamma_{kl, i} + \sum_a \pi^{\phi_a} \phi_{a,i} \, ,
\label{MOMENTUM}
\end{eqnarray}
where $\kappa = 8 \pi G$ and $R$ denotes the Ricci curvature of the 3-metric. 
The object of principal interest is the 
generating functional 
${\cal S} \equiv {\cal S}[\gamma_{ij}(x), \phi_a(x)]$ which assigns
a number (in general complex) to each field configuration 
$\phi_a(x)$ on a space-like hypersurface described by the 
3-metric $\gamma_{ij}(x)$.
The momenta are given by functional derivatives of the generating
functional:
\begin{equation}
\pi^{\phi_a} = {\delta {\cal S} \over \delta \phi_a} \, , \quad 
\pi^{ij} = {\delta {\cal S} \over \delta \gamma_{ij}} \, .
\end{equation}
\end{mathletters}
The constraints equations (\ref{CONSTRAINTS}) are self-contained
equations for the generating functional ${\cal S}$, which is just
the `phase' of the wavefunctional in the semiclassical limit. 

\subsection*{Strong-Coupling Limit: $G \rightarrow \infty$}

To solve for the generating functional, ${\cal S}$, 
one must resort to 
approximation methods. In the limit that Newton's constant $G = \kappa/(8 \pi)$
is large, one may safely neglect the terms appearing within the braces $\{ \}$
of eq.(\ref{ENERGY}). The resulting system, {\it Strongly-Coupled Gravity
and Matter}, is exactly solvable \cite{S98}. Its simplest
solution is of the form
\begin{mathletters}
\begin{equation}
{\cal S} = - 2 \int d^3 x \, \gamma^{1/2} \, H(\phi_a) \, \label{SHE}
\end{equation}
where the function $H \equiv H(\phi_a)$ is a function of the scalar fields
and it satisfies the separated Hamilton-Jacobi equation (SHJE):
\begin{equation}
H^2 = 
{2 \over 3} \, \sum_a \left ( { \partial H \over \partial \phi_a }\right)^2
+ {1\over 3} \, V(\phi_a) \, .  
\end{equation}
\end{mathletters}
Explicit general solutions of the SHJE exist for potentials of the
exponential type
\begin{equation}
V(\phi) = V_0 \, {\rm exp} \left ( \sum_a \, B_a \, \phi_a \right ) \, ;
\label{EXPONENTIAL}
\end{equation}
that is when $\ln V(\phi)$ is linear, $H$ is exactly solvable. 
More complicated potentials are obtained by gluing together 
different exponential potentials together \cite{S92}. 

\section*{Non-Gaussian Adiabatic Fluctuations}

For inflation with two scalar fields, 
non-Gaussian fluctuations arise in a model where one glues continuously 
three exponential potentials 
of the type given by eq.(\ref{EXPONENTIAL}). Such a model is illustrated in
Fig.1. The resulting distribution in large angle temperature anisotropies
is shown in Fig. 2. It is strongly non-Gaussian. 
The impact of similar models on the distribution of galaxies has
been investigated by Fan and Bardeen \cite{FB92} as well as 
Moscardini {\it et al} \cite{M93}.

The challenge in the future is to see if the parameters of the model 
may be adjusted to fit the non-Gaussian signal of Ferreira {\it et al}
\cite{FMG98} and yet be consistent with the limits of Kogut {\it et al} 
\cite{K96}.

\section*{Discussion}

Recent analysis suggests that the COBE DMR map 
may have a non-Gaussian signal \cite{FMG98}. 
It is definitely possible to construct inflation models which produce
non-Gaussian fluctuations \cite{S92}. By studying the statistics of microwave
background fluctuations, one hopes to develop a powerful probe of the very 
early Universe. The situation should improve dramatically
when the next generation of microwave background anisotropy satellites
such as MAP and PLANCK are launched.

\section*{Figure 1}

Non-Gaussian fluctuations 
may be generated during inflation when two scalar fields pass
over a sharp but continuous feature in their potential. 
The light solid curves are lines of uniform scalar field potential;
the heavy lines are the boundaries between the 3 regions where
$\ln V$ is linear.  By choosing the initial conditions, 
the scalar field
trajectories (broken lines) which start in the lower left hand corner
may pass sufficiently near the origin, 
and nonlinear effects become important.

\section*{Figure 2}

For the model described in Fig. 1, a very significant non-Gaussian
histogram in large angle cosmic microwave maps is produced. This 
extreme distribution is 
shown for illustrative purposes. By adjusting the input parameters, one may 
arbitrarily adjust the height and placement of the left peak.
For comparison, the dark curve is the best fit Gaussian. 

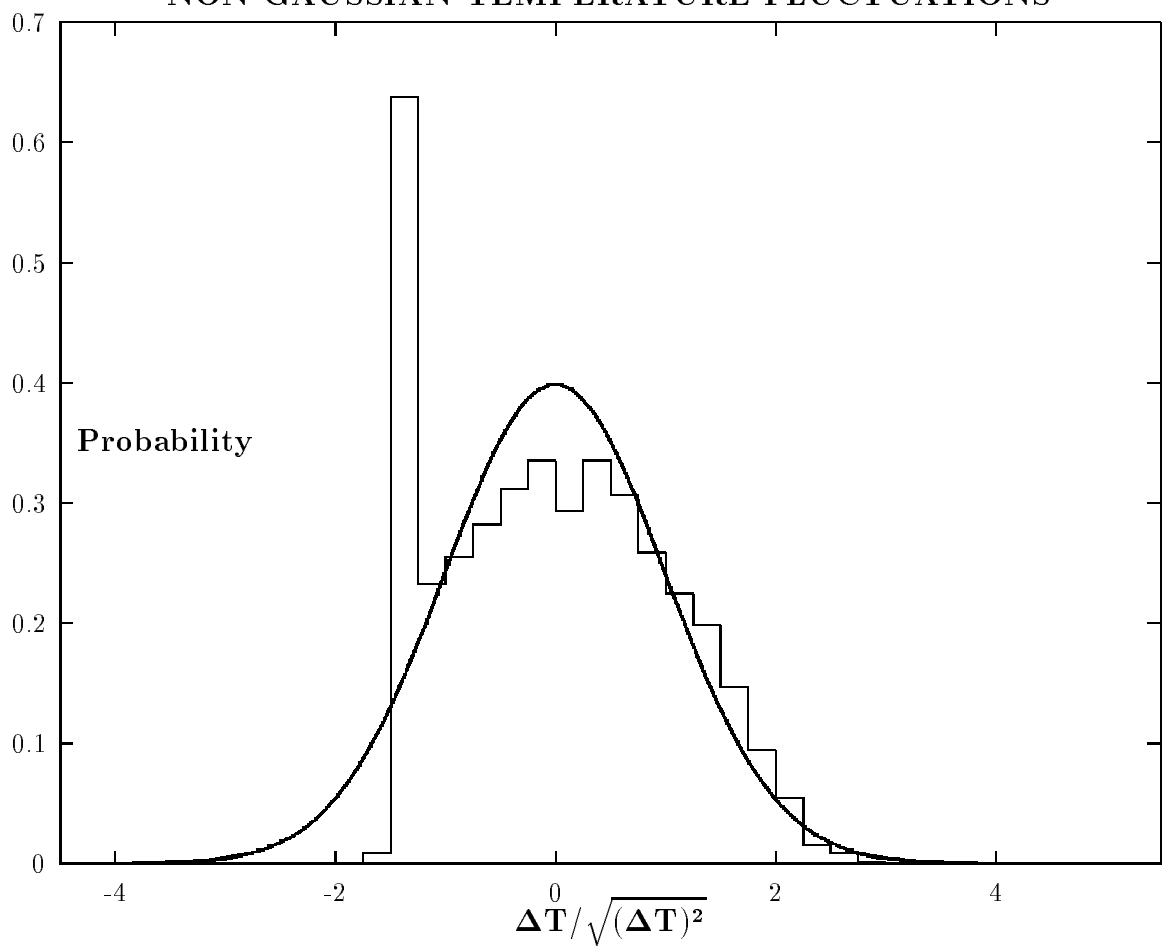
\end{document}